\documentclass{PoS}

\title{Meson spectroscopy with unitary coupled-channels model for heavy-meson decay into three mesons}

\ShortTitle{Meson spectroscopy with unitary coupled-channels model}

\author{\speaker{Satoshi X. Nakamura}\thanks{Present Affiliation: Yukawa
Institute for Theoretical Physics, Kyoto University; nakamura@yukawa.kyoto-u.ac.jp
}\\
Excited Baryon Analysis Center, Jefferson Lab, Newport News, VA, 23606, USA\\
        E-mail: \email{satoshi@jlab.org}}

\abstract{
We develop a model for describing excited mesons decay into three
mesons.
The properties of the excited mesons can be extracted with this model.
The model maintains the three-body unitarity that has been missed in
previous data analyses based on the conventional isobar models.
We study an importance of the three-body unitarity in extracting hadron
properties from data.
For this purpose, we use the unitary and isobar models to analyze the same
pseudo data of $\gamma p \to \pi^+\pi^+\pi^-n$, and extract the
properties of excited mesons. 
We find a significant difference between the unitary and isobar models
in the extracted properties of excited mesons, such as the mass, width and coupling
strength to decay channels.
}

\FullConference{Sixth International Conference on Quarks and Nuclear Physics,\\
		April 16-20, 2012\\
		Ecole Polytechnique, Palaiseau, Paris}

\begin{document}

\section{Introduction}
Hadron properties such as quantum numbers (spin, parity, etc.), mass and
(partial) width have been long studied as a subject called hadron
spectroscopy.
The hadron properties provide important information for understanding
internal structure of the hadron and dynamics which governs it.
The dynamics here is of course QCD in its nonperturbative regime.
The hadron properties can be extracted from data through a careful
analysis, in many cases, partial wave analysis (PWA).
Thus it is essential for hadron spectroscopy to have a reliable theoretical analysis tool.

\begin{figure}[b]
\begin{center}
 \includegraphics[width=\textwidth]{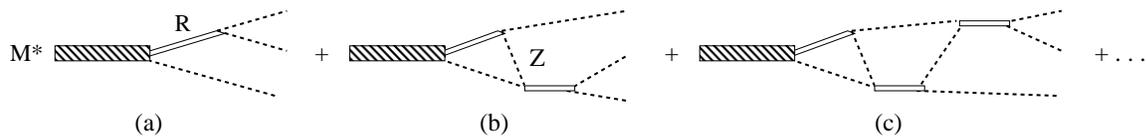}
\end{center}
\caption{\label{fig:mstar-decay3}
$M^*$-decay amplitude.}
\end{figure}

Here, we are interested in analyzing data in which excited mesons decay into three mesons.
Such an analysis has been done, for example, in Ref.~\cite{e852} in
which $\pi N\to M^* N\to \pi\pi\pi N$
($M^*$: an intermediate excited meson) is analyzed to extract the
properties of $M^*$.
The analysis tool which has been conventionally used for this kind of
analysis is the so-called isobar model.
In the isobar model, it is assumed that two mesons form a resonance
($f_0, \rho, f_2$, etc.), and the third meson is treated as a
spectator, as illustrated in Fig.~\ref{fig:mstar-decay3} (a).
A naive question that arises here is how significant the missing three-body unitarity
and coupled-channels effect are.
In order to take care of the three-body unitarity, we need to consider
the Z-diagrams [Fig.~\ref{fig:mstar-decay3} (b)].
Resumming all-order multiple scattering diagrams due to the Z-diagrams,
as shown in Fig.~\ref{fig:mstar-decay3}, 
we obtain a $M^*$ decay amplitude that maintains the three-body unitarity,
and also coupled-channels effects as a consequence.

In this contribution, we would like to address a question how the
three-body unitarity makes a difference in extracting hadron properties
from data~\cite{3pi-short}.
To address this question, we construct both unitary and isobar models,
and fit them to the same data. 
Then we extract and compare $M^*$ properties from the two models.
We will conduct this analysis for the 
$\gamma p \to \pi^+\pi^+\pi^-n$ reaction.
The CLAS Collaboration tried to find an exotic meson
in this reaction using an isobar model~\cite{clas}, and the GlueX Collaboration plans to do a
more elaborate study on this~\cite{gluex}.
To pin down the existence of exotic states, 
a reliable analysis tool is essential.
Therefore, it is interesting to study differences between the unitary and
isobar models for this reaction. 

The organization of the rest of this report is as follows:
We discuss our unitary model in Sec.~\ref{sec:model}.
We also define an isobar model in the same section.
In Sec.~\ref{sec:result},
we determine parameters contained in the unitary model using a
result from a $^3P_0$ model calculation~\cite{3p0}.
Then we use the unitary model to generate pseudo data (Dalitz plots) 
for $\gamma p \to \pi^+\pi^+\pi^-n$, 
and fit the data with the isobar model.
$M^*$ properties are extracted from the two models, and are compared.
%Finally, we give a summary in Sec.~\ref{sec:summary}.

\section{Unitary coupled-channels model}\label{sec:model}

\begin{figure}[t]
 \begin{center}
 \includegraphics[width=50mm]{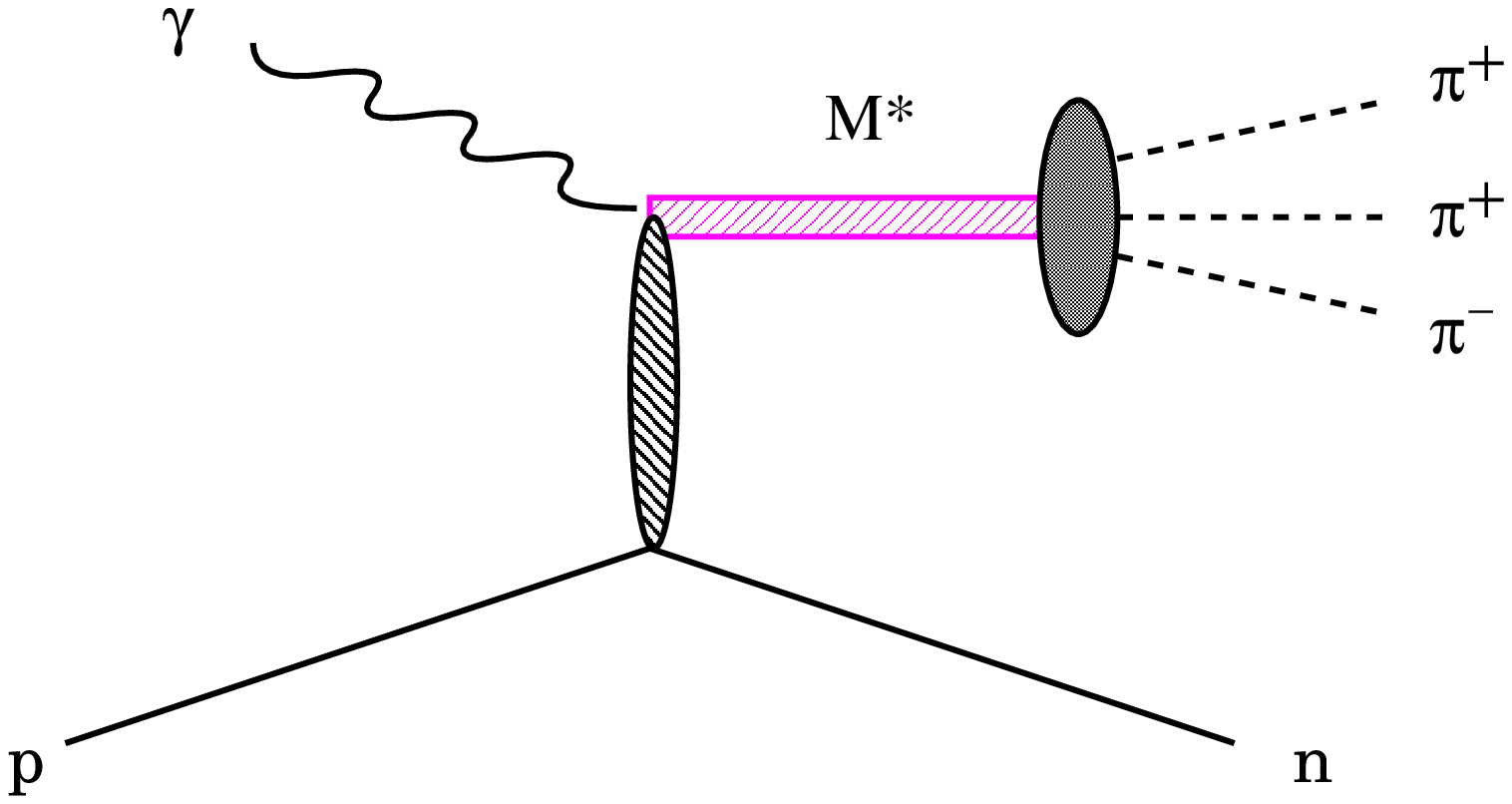}\hspace{10mm}
 \includegraphics[width=90mm]{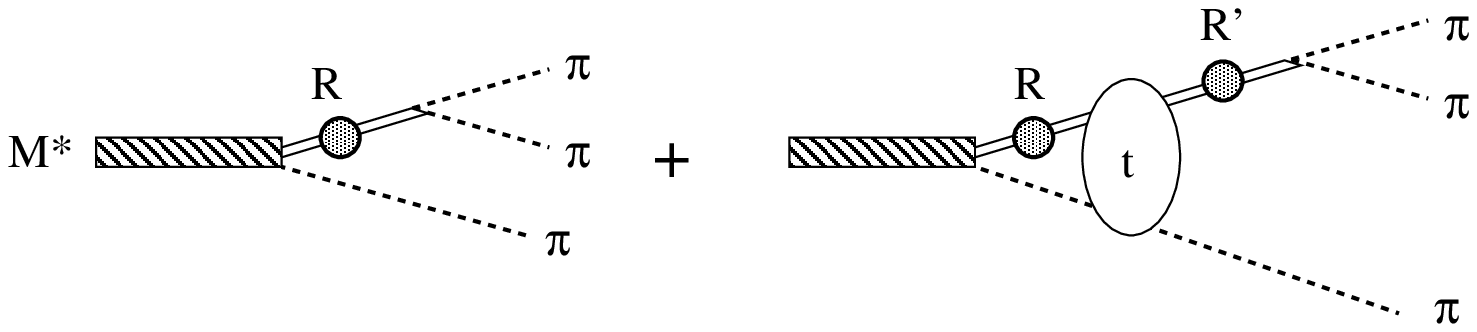}
 \end{center}
 \caption{\label{fig:gp-3pi}
(Left) Schematic $\gamma p \to \pi^+\pi^+\pi^-n$ mechanism considered in
  this work; 
(Middle, Right) $M^*$ decay amplitude calculated with the unitary model. The
  label ``t'' is the $\pi R$ amplitude, and the gray circles near $R$
  stand for intermediate loops of $\pi\pi$ and $K\bar K$.
The isobar model considers only the middle diagram.
 }
 \end{figure}

We assume that the $\gamma p \to \pi^+\pi^+\pi^-n$ reaction proceeds
through $\gamma p \to M^* n$ and $M^*\to \pi^+\pi^+\pi^-$, schematically
shown in Fig.~\ref{fig:gp-3pi} (left). 
There are some other mechanisms, such as an excitation of nucleon
resonances and/or the Deck effect, that lead to the same final state.
We do not consider these backgrounds because we do not deal with actual
experimental data as we mentioned.
For our limited purpose of investigating effects of three-body
unitarity, we expect that these backgrounds would not play an important role.
Thus in the following, we give a discussion on
the $M^*\to 3\pi$ amplitude described with our unitary model 
(a detailed discussion is given in Ref.~\cite{3pi}), 
and then discuss $M^*$ production mechanism and its propagation.
The $\gamma p \to \pi^+\pi^+\pi^-n$ amplitudes are product of these
ingredients, and are used to calculate Dalitz plots distributions of
the three pions.

\vspace{2mm}
\noindent {\bf $M^*$ decay amplitude}\ \ \ 
We assume that an $M^*$ decays through $M^*\to \pi R\to \pi\pi\pi$
(Fig.~\ref{fig:mstar-decay3})
where $R$ stands for a bare state formed by two pions. 
Two pions from $R$-decay can form $R$ again.
Through this type of rescatterings, $R$ is dressed by the pion cloud to
develops a resonance pole corresponding to, for example, $f_0(600), \rho(770)$, etc.
Therefore, we include these resonances in our model as poles in the
$\pi\pi$ scattering
amplitude; we do not use Breit-Wigner functions that has been often used
in previous isobar-model analyses.
Meanwhile, one of two pions from $R$-decay can also interact with the other
pion to form $R$. 
We call this mechanism the $Z$-diagram [Fig.~\ref{fig:mstar-decay3} (b)]. 
Thus a basic ingredient in our model for describing the final state
interaction of $M^*$-decay is 
the $R\leftrightarrow \pi\pi$ interaction. 
This interaction can be fixed by analyzing $\pi\pi$ scattering data. 
We developed a simple coupled-channels ($\pi\pi, K\bar K$) model for the
$\pi\pi$ scattering~\cite{3pi}.
For each partial wave,
$\pi\pi\ ({\rm or}\ K\bar K) \to \pi\pi\ ({\rm or}\ K\bar K)$ potentials are given by
$s$-channel exchange of bare $R$ states,
and the partial wave amplitude is obtained by solving three-dimensional
Lippmann-Schwinger equation.
We obtained a reasonable description of empirical $\pi\pi$ phase shifts
and inelasticities for 
$(L,I)=(0,0),\ (1,1),\ (2,0)$ partial waves
($L$: orbital angular momentum; $I$: total isospin)
from the threshold to 2~GeV. 
We extracted pole positions from the partial wave amplitudes, and found
that they are in good agreement with the listing in the Particle Data Group.

Having fixed the $R\leftrightarrow \pi\pi$ vertices,
we now have a coupled-channels scattering equation for $\pi R$ amplitudes,
and the Z-diagrams work as the driving force.
Different $\pi R$ channels are coupled via the Z-diagrams.
The $R$ Green function contains self energies from the $\pi\pi$ and
$K\bar K$ loops.
Both Z-diagrams and $\pi R$ Green functions contain three-pion unitarity
cut, and the resultant three-pion amplitude (from the $\pi R$ amplitude) 
satisfies the three-body unitarity.
The three-pion unitarity cut in the Z-diagrams makes it difficult to
solve the $\pi R$ scattering equation with the standard subtraction
technique. 
We handle the problem with the Spline method which is discussed in
detail in Ref.~\cite{msl}.

The $\pi R$ amplitudes are used to calculate the $M^*$ decay amplitudes,
as seen in Fig.~\ref{fig:gp-3pi}(middle,right), by the convolution with
$M^*\to \pi R$ vertices.
By retaining only the middle diagram of Fig.~\ref{fig:gp-3pi}, the $M^*$
decay amplitude is similar to those used in the isobar models. Thus we
define our isobar model by considering only this mechanism.

\vspace{2mm}
\noindent {\bf Green function and production mechanism of $M^*$}\ \ \ 
The $M^*$ Green function for the unitary model is given by 
$G^{-1}(W)=W-m^0_{M^*}-\Sigma(W)$, where $W$ is the total energy of the
three pions in their CM frame, 
$m^0_{M^*}$ is the bare mass of $M^*$, and $\Sigma(W)$ is the self
energy.
The self energy contains the $\pi R$ amplitudes, so that the self energy
has a consistency with the $M^*$ decay amplitude 
(see Ref.~\cite{3pi} for expressions). 
The three-body unitarity requires this consistency.
Pole positions corresponding to $M^*$ resonances are a quantity of
interest. 
We search for pole positions ($M_R$) that satisfies $G^{-1}(M_R)=0$.
The pole search needs an analytic continuation of the amplitudes to the
complex energy plane. 
A method of the analytic continuation suitable for our unitary model has
been developed in Ref.~\cite{suzuki}.
For the isobar model, on the other hand, we use a Breit-Wigner function,
as has been done in most isobar-model analyses.

For the $M^*$ production mechanism,
we assume a $t$-channel pion exchange, and the pion interacts with 
a $\rho$-meson from the incident photon, invoking the vector-meson
dominance. 
Although this is a rather simple assumption and often not very
realistic, this model still works enough for our purpose of studying
effects of the three-body unitarity.
We use this $M^*$ production mechanism for both of the unitary and
isobar models.

\section{Numerical results}\label{sec:result}

Having described our unitary and isobar models, we now use them for a
numerical analysis.
As mentioned in the introduction, we want to learn how three-body
unitarity makes a difference in extracting $M^*$ properties from Dalitz
plot analysis.
For this purpose, we take the following procedure:
(i) We determine $M^*\to \pi R$ couplings for the unitary model using a prediction from the
$^3P_0$ model for partial width for $M^*\to \pi R$ decays~\cite{3p0};
(ii) We generate pseudo data with the unitary model;
(iii) We fit the data with the isobar model;
(iv) We extract the $M^*$ properties from the two models.
Details are discussed in the following paragraphs.

\begin{figure}[b]
\begin{center}
 \includegraphics[width=75mm]{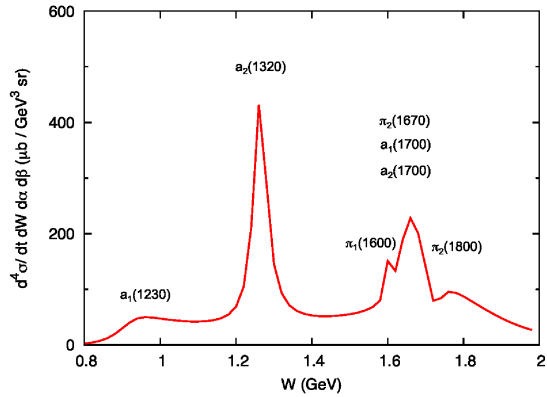}
\end{center}
\caption{\label{fig:w-dep}
$W$-dependence of integrated Dalitz plots from the unitary model.
The $M^*\to \pi R$ couplings in the model have been fixed using the
 $^3P_0$ model prediction.
$M$'s contributing to the peaks are indicated.
}
\end{figure}

\begin{figure}[t]
\begin{center}
 \includegraphics[width=75mm]{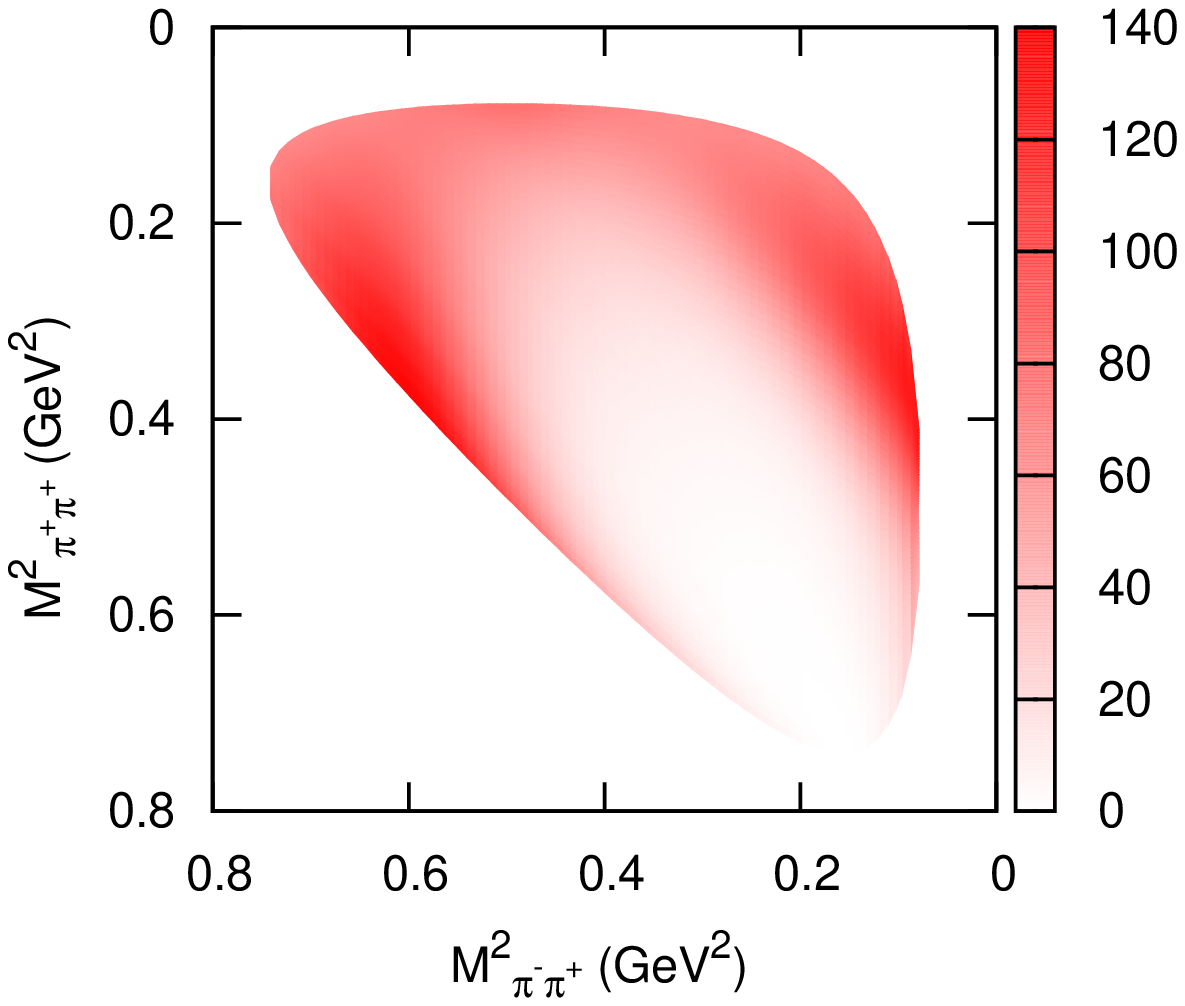}
 \includegraphics[width=75mm]{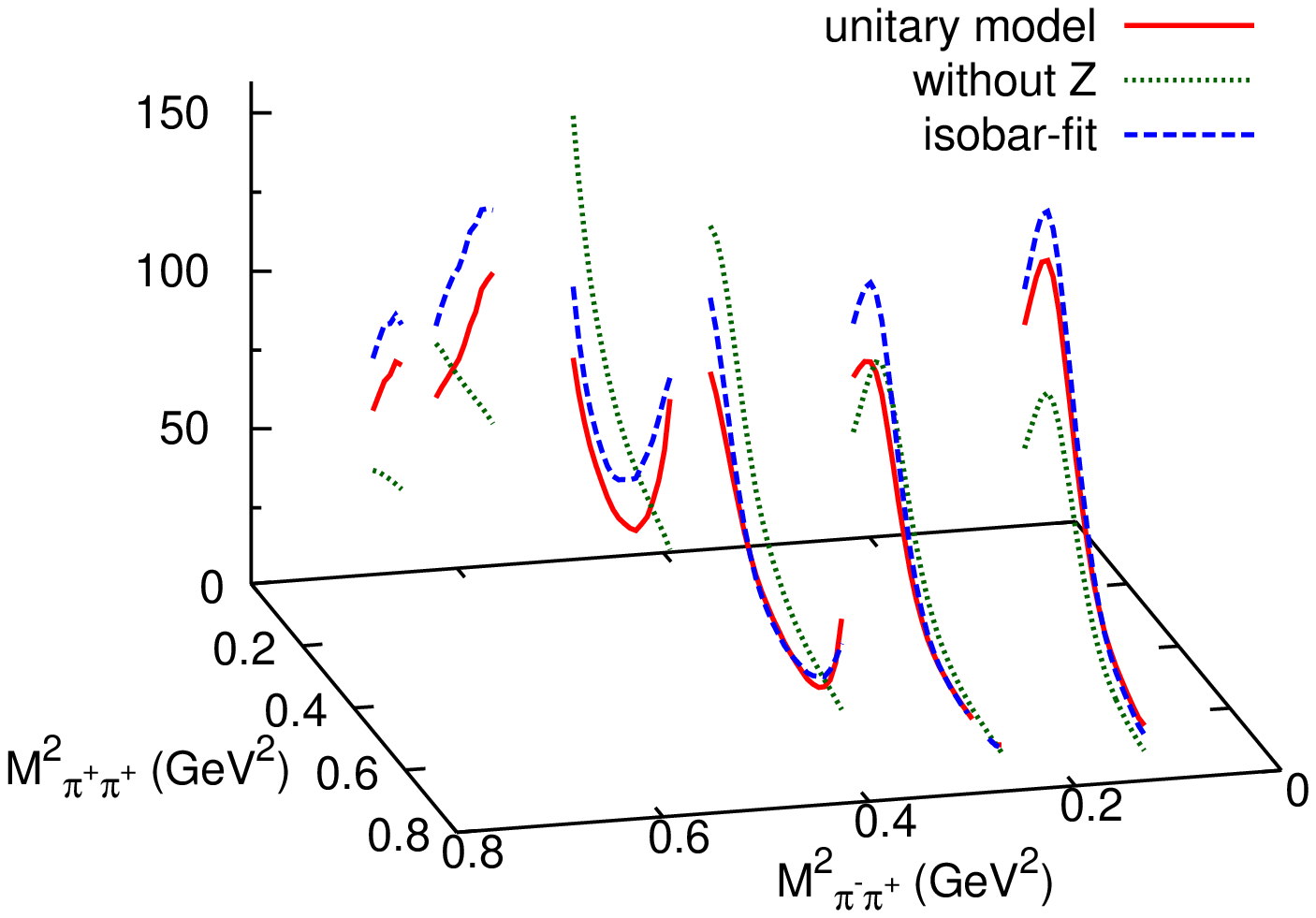}
\end{center}
\caption{\label{fig:dalitz} (Left) Dalitz plot distribution of 
$\pi^+\pi^+\pi^-$ in the 
$\gamma p \to \pi^+\pi^+\pi^-n$ reaction at $W=1$~GeV.
(Right) 3-dimensional view of the Dalitz plot distribution shown on the left.}
\end{figure}

We consider partial waves and $M^*$s that have been included in the CLAS
analysis of $\gamma p \to \pi^+\pi^+\pi^-n$~\cite{clas}.
They are $J^{PC}=1^{++}$ [$a_1(1230)$, $a_1(1700)$],
$2^{++}$ [$a_2(1320)$, $a_2(1700)$], 
$2^{-+}$ [$\pi_2(1670)$, $\pi_2(1800)$], and 
$1^{-+}$ [$\pi_1(1600)$].
The $^3P_0$ model, a quark model, can predict partial widths for
$M^*\to \pi R$ decays~\cite{3p0}.
We use the prediction to fix $M^*\to \pi R$ coupling constants, assuming
that the phases are $+1$.
% and the cutoffs are set to 1~GeV.
%
With the parameters fixed in this way, we run the unitary model to
generate Dalitz plots.
We chose the following kinematics:
$E_\gamma=$5~GeV 
($E_\gamma$: incident photon energy in the laboratory frame);
$t=-0.4$~GeV$^2$ ($t$: squared four-momentum transfer of the nucleon);
0.8~GeV $\le W\le$ 2~GeV.
Finally, three-pion orientation is chosen so that the plane formed by
the three pions is perpendicular to the final nucleon momentum in the
three-pion CM frame.
In Fig.~\ref{fig:w-dep}, we show a spectrum obtained by integrating the
three pion distribution within the specified kinematics.
As seen in the figure, different $M^*$s play major roles at
different $W$ regions, depending on their masses and widths.
At each $W$, the final three pions are distributed in a pattern
characteristic to the associated $M^*$.
For example, in Fig.~\ref{fig:dalitz} (left),
we show a Dalitz plot at $W=$1~GeV where 
the main contribution is from the $a_1(1230)$ decay.
A three-dimensional view of the same Dalitz plot is also shown by the solid
curve in Fig.~\ref{fig:dalitz} (right).
On the same figure, the dotted curve is obtained by 
simply turning off the Z-diagrams from the $M^*$ decay
amplitudes.
Thus the difference between the solid and dotted curves is the effect of
the Z-diagrams (of the $M^*$ decay amplitude)
on the Dalitz plot.
The Dalitz plot distributions are fitted with the isobar model.
Fitting parameters are (complex) $M^*\to \pi R$ couplings and cutoffs,
the Breit-Wigner masses and widths in the $M^*$ Green functions.
The Breit-Wigner width also contains an adjustable parameter to fit the
energy dependence.
We also include a flat (in three-pion distribution), non-interfering
background term, as usually done in the isobar-model analyses.
The strength of the flat background is also a fitting parameter. 
The quality of the fit can be seen by comparing the dashed curve (fit)
with the solid curve on the right panel of Fig.~\ref{fig:dalitz}.
The isobar model reasonably fits the Dalitz plots from the unitary model.

\begin{figure}[t]
\begin{center}
 \includegraphics[width=60mm]{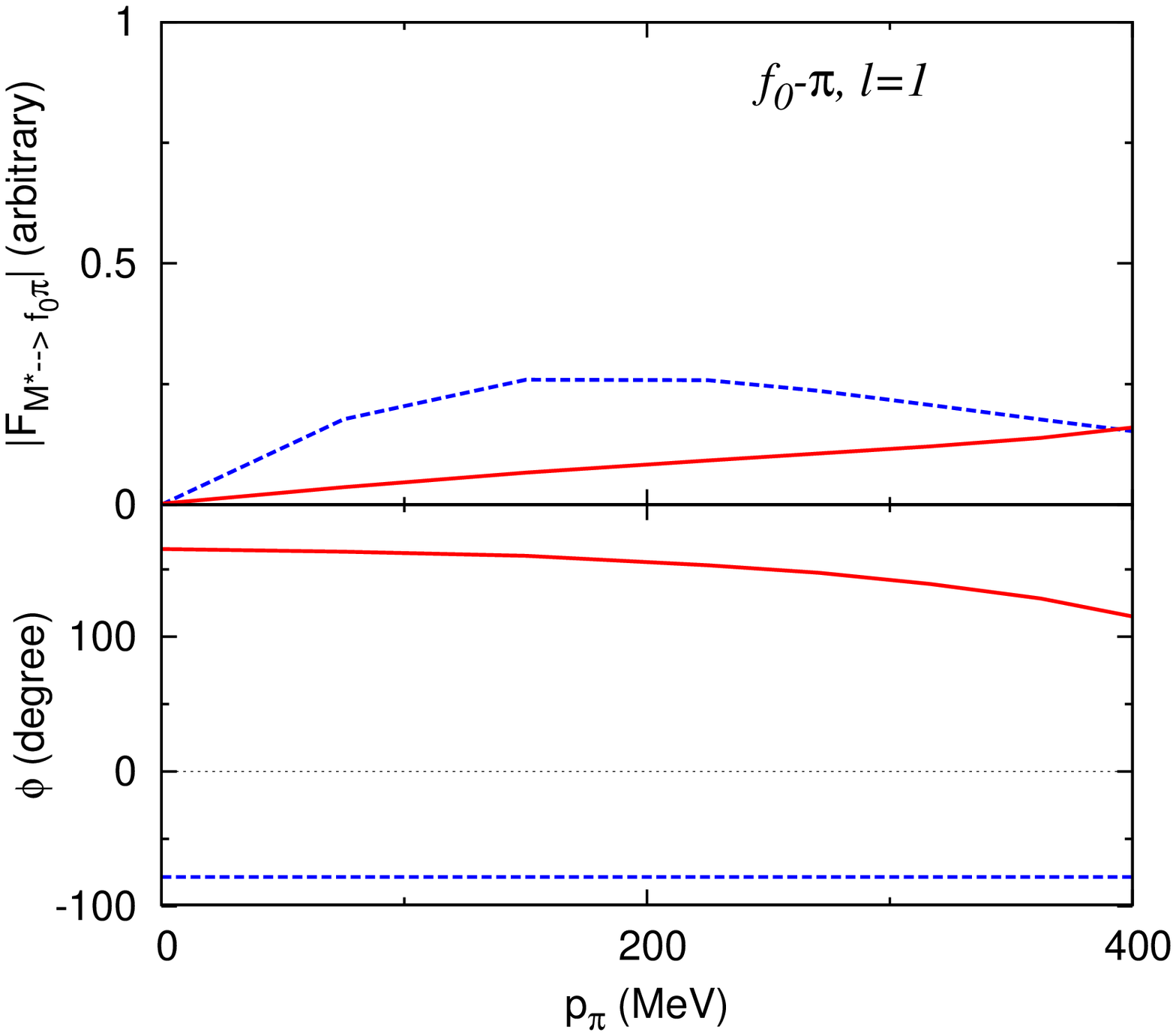}
\hspace{12mm}
 \includegraphics[width=60mm]{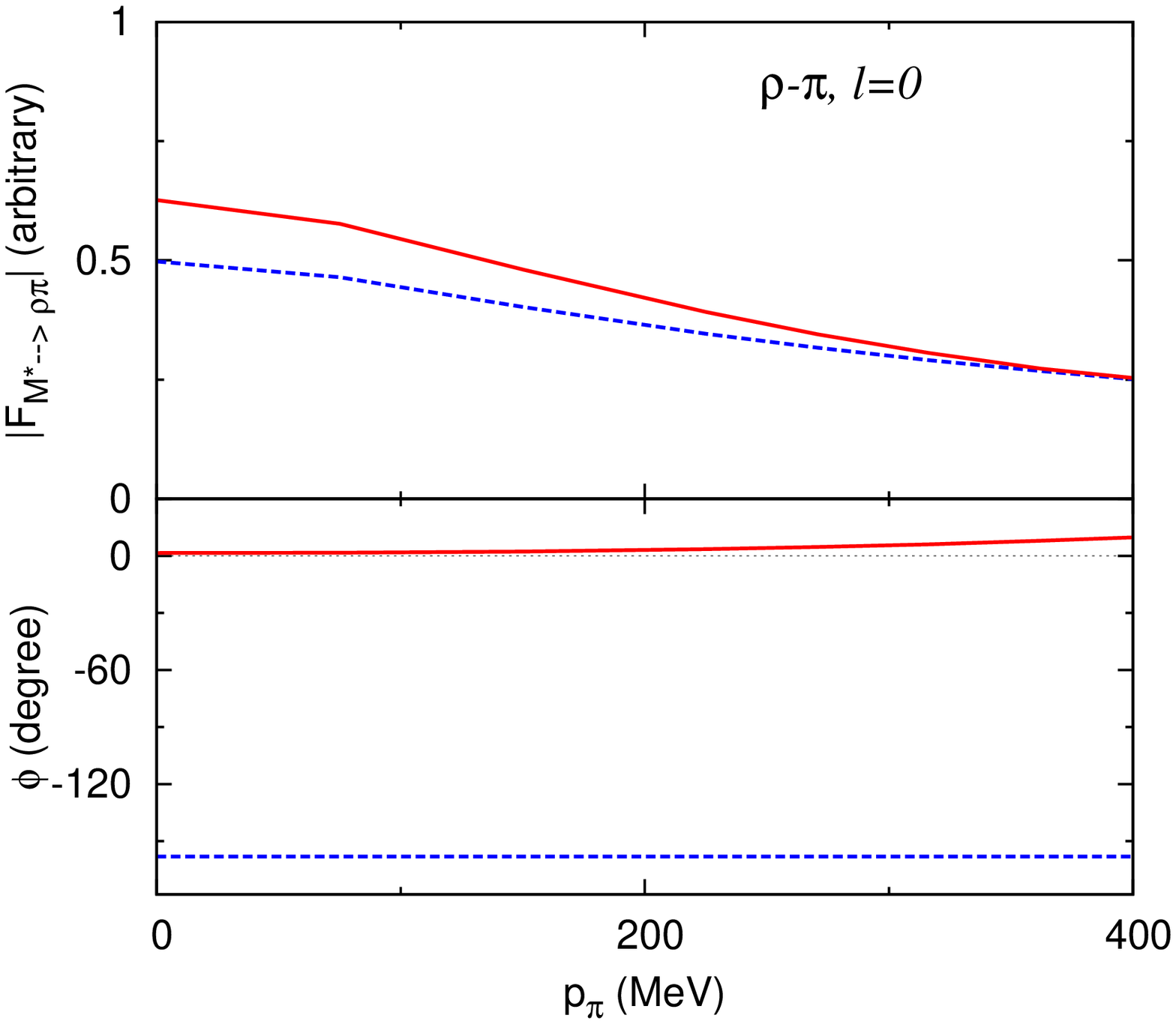}
\end{center}
\caption{\label{fig:vertex}
The strength (phase) of $M^*\to \pi R$ vertices on the upper
 (lower) half. 
The unitary (isobar) model is displayed by the solid (dashed) lines.
(Left) $a_1$ (1230) $\to \pi f_0$ vertex in $l=1$ 
($l$: orbital angular momentum between $\pi$ and $R$). 
(Right) $a_1$ (1230) $\to \pi \rho$ vertex in $l=0$.
}
\end{figure}

We now discuss the $M^*$ properties extracted from the unitary and
isobar models.
The mass (width) of an $M^*$ of the unitary model is defined with the
pole position, $M_R$, by ${\rm Re} [M_R]$ ($-2\;{\rm Im} [M_R]$).
These are compared with Breit-Wigner
masses and widths from the isobar model in Table~\ref{tab:pole}.
While we can see a good agreement for narrow $M^*$
[$a_2(1320)$, $\pi_1(1600)$], there are also significant differences
for broader $M^*$ [$a_1(1230)$, $\pi_2(1670)$].
\begin{table}[t]
\caption{\label{tab:pole} $M^*$ masses and widths from
the unitary (UT) and the isobar (IB) models.
}
% \begin{center}
%\hspace{3mm}
\renewcommand{\arraystretch}{1.1}
\tabcolsep=2.5mm
 \begin{tabular}[t]{c|ccccccccccc}\hline
&\multicolumn{2}{c}{$a_1(1230)$}&
\multicolumn{2}{c}{$a_2(1320)$}&
\multicolumn{2}{c}{$\pi_1(1600)$}&
\multicolumn{2}{c}{$\pi_2(1670)$}
&\multicolumn{2}{c}{$a_1(1700)$}
\\
& UT& IB&UT& IB&UT& IB&UT& IB&UT& IB \\\hline
Mass (MeV) &  937 & 1096 &1263 & 1269 &1599 & 1599 &1784 & 1798 & 1658  &  1687 \\
Width (MeV)&  496 &  684 &42   &  56  &8    &    8 &456  &  502 &  106  &   99
\\\hline
 \end{tabular}
%%
% \begin{tabular}[t]{c|cccccccc}
%&\multicolumn{2}{c}{$a_1(1230)$}&
%\multicolumn{2}{c}{$a_2(1320)$}&
%\multicolumn{2}{c}{$\pi_1(1600)$}&
%\multicolumn{2}{c}{$\pi_2(1670)$}
%\\
%& UT& IB&UT& IB&UT& IB&UT& IB \\\hline
%Mass (MeV) &  913 &  &1263 &  &1599 &  &1722 &     \\
%Width (MeV)&  138 &  &42   &  &8    &  &52   &     
%%\\\hline\hline
% \end{tabular}
%%
% \begin{tabular}[t]{c|cccccc}
%&\multicolumn{2}{c}{$a_1(1700)$}&
%\multicolumn{2}{c}{$a_2(1700)$}&
%\multicolumn{2}{c}{$\pi_2(1800)$} \\
%& UT& IB&UT& IB&UT& IB \\\hline
%Mass (MeV) & 1658  & &1652  &  &1784 &     \\
%Width (MeV)&  106  & &  76  &  & 456 &     \\
% \end{tabular}
%\end{center}
\end{table}
Another interesting quantity is the coupling strength of $M^*$ to each
decay channel.
Our result for $a_1(1230)\to \pi R$ vertices are shown in
Fig.~\ref{fig:vertex} as a function of the pion momentum, and the unitary and isobar models are compared.
The difference is significant in both the strength and the phase. 
These results indicate the importance of considering the three-body
unitarity in extracting the $M^*$ properties.

%\section{Summary}\label{sec:summary}

\vspace{5mm}
The author thanks H. Kamano, T.-S. H. Lee and T. Sato for collaboration.
This work is supported by the U.S. Department of Energy, Office of Nuclear Physics
Division, under Contract No. DE-AC05-06OR23177 under which Jefferson Science
Associates operates Jefferson Lab.


\begin{thebibliography}{99}
\bibitem{e852}
S. U. Chung et al. (E852 Collaboration), Phys. Rev. D {\bf 65} (2002) 072001.

\bibitem{3pi-short}
S.X. Nakamura, H. Kamano, T.S.H. Lee and T. Sato, in preparation.

\bibitem{3p0}
T. Barnes, F. E. Close, P. R. Page, and E. S. Swanson, 
Phys. Rev. D {\bf 55}  (1997) 4157.

\bibitem{clas} M. Nozar et al. (CLAS Collaboration), Phys. Rev. Lett.
{\bf 102} (2009) 102002.

\bibitem{gluex}
Presentation to PAC30 (The GlueX Collaboration), GlueX-doc-1226-v1.

\bibitem{msl}
A. Matsuyama, T. Sato, and T.-S. H. Lee, Phys. Rep. {\bf 439}, 193 (2007).

\bibitem{3pi} 
H. Kamano, S.X. Nakamura, T.S.H. Lee and T. Sato, Phys.Rev. D {\bf 84} (2011) 114019.

\bibitem{suzuki} 
N. Suzuki, T.~Sato and T.~S.~H. Lee, Phys. Rev. C {\bf 79} (2009) 025205.



\end{thebibliography}
\end{document}